\documentclass[12pt]{article}
\usepackage{amsmath,amsfonts,amssymb}

\usepackage{indentfirst}
\makeatletter
\renewcommand{\@begintheorem}[2]{\begin{trivlist}
\item[\hspace{\labelsep}{\bfseries#1\ #2.}]\itshape}
\renewcommand{\@opargbegintheorem}[3]{\begin{trivlist}
\item[\hspace{\labelsep}{\bfseries#1\ #2\ (#3).}]\itshape}
\renewcommand{\@endtheorem}{\end{trivlist}}
\makeatother
\author{Sergey V.\,Smirnov\thanks{Department of Mathematics and Mechanics, Moscow State University. E-mail: {\tt ssmirnov@higeom.math.msu.su}}}
\title{On Darboux integrability of discrete 2D Toda lattices}

\def\pa{\partial}

\def\phi{\varphi}

\newtheorem{proposition}{\sc Proposition}
\newtheorem{theorem}{\sc Theorem}
\newtheorem{remark}{\sc Remark}
\newtheorem{example}{\sc Example}
\newtheorem{corollary}{\sc Corollary}
\newtheorem{definition}{\sc Definition}

\begin{document}

\maketitle
\begin{abstract}
Darboux integrability of semidiscrete and discrete 2D Toda lattices corresponding to Lie algebras of $A$ and $C$ series is proved.
\end{abstract}

\section{Introduction}

Interest to the problem of discretization of well-known integrable systems appeared and then started to grow after rapid development of theory
of integrable systems in 70-80-ies of the previous century. Due to the fact that now there are several dominating (but not equivalent)
approaches to integrability, when an integrable system is discretized, usually one of the definitions is being chosen and then attempts to
construct a discrete system possessing this property and tending to initial continuous analog in the continuum limit are carried out. Properties
like existence of sufficient amount of conservation laws or first integrals, existence of symmetries, possibility to express an equation in a
Lax form or existence of multi-soliton solutions are considered as integrability criteria.

In the theory of hyperbolic equations of the form
\begin{equation}
\label{hypeqn}
u_{xy}=F(x,y,u,u_x,u_y)
\end{equation}
integrals in directions $x$ and $y$, i.e. functions depending on dynamic variables such that their total derivative with respect to variables $x$ and
$y$ correspondingly vanish on solutions of the equation~(\ref{hypeqn}), are natural analogs of first integrals of ordinary differential equations.
However there is principal difference between one-dimensional and two-dimensional cases. In one-dimensional case each ordinary differential
equation has first integrals (although it could be very difficult to find them and possibly these integrals can not be expressed in quadratures).
But in two-dimensional case existence of $x$- or $y$-integrals is exceptional.

In this paper we study semidiscrete and purely discrete analogs of the so-called {\it generalized two-dimensional Toda lattices} having sufficiently
many independent integrals in both directions (i.e. sufficiently many $n$-integrals and $x$-integrals in the semidiscrete case and sufficiently many
$n$-integrals and $m$-integrals in the purely discrete case).

Various Toda chains and lattices are being examined already for a considerable period of time. One-dimensional Toda chain that describes a system of particles
on a line with exponential interaction between each pair of neighboring particles was introduced
by M.\,Toda in 1967 in his paper~\cite{To}. After that Bogoyavlensky~\cite{Bo} introduced generalized one-dimensional Toda chains corresponding to simple Lie
algebras in 1976. In the end of 1970-ties and in the beginning of 1980-ties a series of papers (see~\cite{Mi}--\cite{LSSh}) where
generalized two-dimensional Toda lattices were considered appeared almost at the same time. In 1991 Suris~\cite{Su} studied discrete generalized one-dimensional
Toda chains.

In 1981 Shabat and Yamilov~\cite{ShJa} have considered the so-called {\it exponential systems}, i.e. systems of hyperbolic partial differential equations
of the form
\begin{equation}
\label{expsyst}
u_{xy}(j)=\exp\left(\sum\limits_{k=1}^r a_{jk} u(k)\right),\quad j=1,2,\dots, r,
\end{equation}
where $a_{jk}$ are constants and functions $u(j)$ depend on variables $x$ and $y$. It is easy to verify that if the matrix $K=(a_{jk})$ is the Cartan matrix of a
simple Lie algebra of $A$-series, than the system~(\ref{expsyst}) is equivalent to the two-dimensional Toda lattice
\begin{equation}
\label{qt}
q_{xy}(j)=\exp(q(j+1)-q(j))-\exp(q(j)-q(j-1)),\quad j=0,1,\dots,r,
\end{equation}
with trivial boundary conditions $q(-1)\equiv\infty$, $q(r)\equiv-\infty$. Exponential systems corresponding to simple Lie algebras of other series are sometimes
called {\it generalized Toda lattices}. In papers~\cite{ShJa,LSSh} the notion of {\it characteristic algebra} for the system of the form~(\ref{expsyst})
was introduced and it was proved that the charactiristic algebra of an irreducible exponential system is finite-dimensional if and only if the corresponding matrix $K$
is the Cartan matrix of a simple Lie algebra. And finite dimensionality of characteristic algebra, in turn, is equivalent to the existence of a complete family of
$x$- and $y$-integrals.

In the case $r=1$ the Toda lattice~(\ref{qt}) is known to be equivalent to the Liouville equation that has been studied already in 1853~\cite{Liouv}. Semidiscrete
and discrete analogs of the Liouville equation were introduced and examined by Adler and Startsev in 1999~\cite{AS}. Besides this, other attempts to construct
and to study discrete analogs of exponential systems were made. Lax representations for purely discrete analogs of the $A$-series and $C$-series Toda lattices were
obtained in~\cite{Wa} and in~\cite{Ha06} respectively. Semidiscrete analog of $C$-series Toda lattices was examined in~\cite{Sm}.

Systematic approach, based on the idea that ``correct'' discretization has the same integrals that its continuous analog, was used by Habibullin, Zheltukhin
and Yangubaeva in~\cite{HZY} to construct semidiscrete analogs of the exponential systems. After that the same approach was applied in~\cite{GHY} to obtain purely
discrete analogs of generalized Toda lattices. In both cases all discrete Toda lattices, that were known before, appeared to be particular cases of
these (semi)discrete exponential systems.

The paper is structured as follows. Two rather simple constructions that allow to obtain $x$- and $y$-integrals for $A$-series Toda lattice and its reductions in
the continuous case are described in the Section~\ref{cont}. The first approach is based on the use of nonlocal variables, introduced by Shabat in~\cite{Sh95b},
and the second one uses Darboux--Laplace transformations. Similar constructions for the semidiscrete Toda lattices are developed in the Section~\ref{semidisc};
complete families of independent $x$- and $n$-integrals for $A$- and $C$-series Toda lattices are obtained. Section~\ref{disc} is devoted to construction of complete
families of $m$- and $n$-integrals for $A$- and $C$-series Toda lattices in the purely discrete case.

\section{Continuous case}\label{cont}
\subsection{Darboux--Laplace transformations}

Interest to Toda lattices have arisen in mathematical physics only in the second half of the 20-th century. However the two-dimensional Toda lattice had appeared
in problems of classical differential geometry long before this. Sequences of Darboux--Laplace transformations leading to the Toda lattice equations were
used in the 19-th in the theory of conjugate coordinate nets (see~\cite{Da}). However we won't go deep into differential-geometric nature of the Toda lattice, we'll only describe
the construction that leads to the corresponding equations.

Two linear hyperbolic differential operators
$$
{\mathcal L}=\pa_x\pa_y+a\pa_x+b\pa_y+c\quad\hbox{and}\quad\hat{\mathcal L}=\pa_x\pa_y+\hat a\pa_x+\hat b\pa_y+\hat c
$$
in two variables are related by a {\it Darboux--Laplace transformation}, if there exist first order operators ${\mathcal D}=\alpha\pa_x+\beta\pa_y+\gamma$ and
$\hat{\mathcal D}=\hat\alpha\pa_x+\hat\beta\pa_y+\hat\gamma$ such that the following relation is satisfied:
\begin{equation}
\label{dlt}
\hat{\mathcal L}{\mathcal D}=\hat{\mathcal D}{\mathcal L}.
\end{equation}
It's easy to verify that the operator ${\mathcal L}$ can be rewritten in the following form:
$$
{\mathcal L}=(\pa_x+b)(\pa_y+a)+k=(\pa_y+a)(\pa_x+b)+h,
$$
where functions $k$ and $h$ are defined by the equations $k=c-ab-a_x$ and $h=c-ab-b_y$. These functions are called {\it the Laplace invariants} of the operator
${\mathcal L}$\footnote{Functions $k$ and $h$ are called {\it invariants} because they are invariant under gauge transformations
${\mathcal L}\mapsto\bar{\mathcal L}=\omega^{-1}{\mathcal L}\omega$, where $\omega=\omega(x,y)$.}.
The operator relation~(\ref{dlt}) is equivalent to the following system of equations on coefficients of the operators:
\begin{eqnarray}
\label{eee}
\left\lbrace
\begin{array}{l}
\alpha-\hat\alpha=0\\
\beta-\hat\beta=0\\
\alpha_y+\alpha\hat a-\hat\alpha a=0\\
\beta_x+\beta\hat b-\hat\beta b=0\\
\alpha_x+\beta_y+\alpha\hat b-\hat\alpha b+\beta\hat a-\hat\beta a+\gamma-\hat\gamma=0\\
\hat{\mathcal L}(\alpha)-\hat\alpha(c+a_x)-\hat\beta a_y+\gamma_y+\hat a\gamma-\hat\gamma a=0\\
\hat{\mathcal L}(\beta)-\hat\beta(c+b_y)-\hat\alpha b_x+\gamma_x+\hat b\gamma-\hat\gamma b=0\\
\hat{\mathcal L}(\gamma)-\hat\alpha c_x-\hat\beta c_y-\hat\gamma c=0
\end{array}
\right..
\end{eqnarray}
Complete description of all transformation operators ${\mathcal D}$ for a given hyperbolic operator ${\mathcal L}$, i.e. complete solution of the
system~(\ref{eee}), is an open problem (see~\cite{Sh95a}). We are going to examine a particular solution of this problem: we'll consider transformation
operators of the form ${\mathcal D}=\pa_x+\gamma$. In this case the system~(\ref{eee}) leads to the equations
\begin{equation}
\label{paa}
\hat k=h,\quad\hat h=2h-k+(\ln h)_{xy},
\end{equation}
where $\hat k$ and $\hat h$ are the Laplace invariants of the operator $\hat{\mathcal L}$. The system~(\ref{paa}) dates back to Darboux~\cite{Da}.

Consider a sequence of hyperbolic operators $\dots,{\mathcal L}_{j-1},{\mathcal L}_j,{\mathcal L}_{j+1},\dots$, such that any two neighboring operators are
related by a Darboux--Laplace transformation of the type specified in the previous paragraph. Hence the corresponding system of equations can be rewritten
as follows:
\begin{equation}
\label{htoda}
(\ln h(j))_{xy}=h(j-1)-2h(j)+h(j+1),
\end{equation}
where $h(j)$ and $k(j)$ are the Laplace invariants of the operator ${\mathcal L}_j$, and $k(j)=h(j-1)$ for all $j$. One can easily verify that use of
the substitution $h(j)=\exp(q(j+1)-q(j))$ allows to rewrite the equation~(\ref{htoda}) in the form
\begin{equation}
\label{qtoda}
q_{xy}(j)=\exp(q(j+1)-q(j))-\exp(q(j)-q(j-1)),
\end{equation}
and that use of the substitution $h(j)=u_{xy}(j)$ leads to the following equation:
\begin{equation}
\label{utoda}
u_{xy}(j)=\exp(u(j-1)-2u(j)+u(j+1)).
\end{equation}
It's easy to notice that the system~(\ref{utoda}) (or, more precisely, its closure with the boundary conditions $u(-1)=u(r)=-\infty$ for any natural $r$)
is a particular case of the exponential system~(\ref{expsyst}) corresponding to the Cartan matrix of an $A$-series Lie algebra. Equations~(\ref{htoda})--(\ref{utoda})
are various forms of {\it the two-dimensional Toda lattice}.

Consider finite Toda lattices, that is, exponential systems~(\ref{expsyst}), where $K$ is the Cartan matrix of one of the infinite series $A$--$D$ of simple Lie
algebras (we won't consider lattices corresponding to exceptional Lie algebras here). Each of these systems is obtained by imposing of certain boundary conditions
on the infinite Toda lattice~(\ref{qtoda}). The explicit forms of this cut-off constraints are listed below. An $A$-series lattice is defined by the trivial
boundary conditions
\begin{equation}
\label{aseries}
q(-1)=\infty, \quad q(r+1)=-\infty
\end{equation}
for some $r\in\mathbb N$. $B$-series lattices are defined the following cut-off constraint:
\begin{equation}
\label{bseries}
q(0)=0, \quad q(r+1)=-\infty
\end{equation}
for some $r\in\mathbb N$. $C$-series lattices are generated by a cut-off constraints of the form
\begin{equation}
\label{cseries}
q(0)=-q(1), \quad q(r+1)=-\infty,
\end{equation}
and $D$-series lattices are generated by a more complicate cut-off constraint:
\begin{equation}
\label{dseries}
q(0)=-\ln\left(e^{q(2)}-\frac{q_x(1)q_y(1)}{2\sinh q(1)}\right), \quad q(r+1)=-\infty.
\end{equation}
In one-dimensional case the cut-off constraint~(\ref{dseries}) for series $D$ was found in~\cite{AH} and in two-dimensional case it was found in~\cite{GH}.

\subsection{Nonlocal variables}\label{contnonloc}

Symmetries of the infinite two-dimensional Toda lattice cannot be expressed in terms of the dynamic variables (on the contrary to the one-dimensional
case). In the two-dimensional case one has to introduce nonlocal variables in order to define symmetries of the infinite Toda lattice.

Denote $q_x(j)$ by $b(j)$. Hence the Toda lattice can be rewritten as follows:
\begin{equation}
\label{btoda}
b_y(j)=h(j)-h(j-1).
\end{equation}
Define the nonlocal variables $b^{(1)}(j)$ by the identities
\begin{equation}
\label{all}
\pa_x b (j)=b^{(1)}(j)-b^{(1)}(j-1),\quad\pa_y b^{(1)} (j)=\pa_x h(j).
\end{equation}
Compatibility of these equations is provided by the Toda lattice~(\ref{btoda}):
$$
\pa_y\pa_x b (j)=\pa_y \left(b^{(1)}(j)-b^{(1)}(j-1)\right)=\pa_x \left(h(j)-h(j-1)\right)=\pa_x\pa_y b(j).
$$
Further on, it is necessary to determine the derivatives $\pa_x b^{(1)}(j)$, but this, however, requires to introduce nonlocalities $b^{(2)} (j)$
of higher order. Set
$$
\pa_x b^{(1)} (j)=b^{(1)}(j)\left(b(j+1)-b(j)\right)+b^{(2)}(j)-b^{(2)}(j-1);
$$
hence the compatibility condition $\pa_y\pa_x b^{(1)}(j)=\pa_x\pa_y b^{(1)}(j)$ obviously leads to the following identity:
$$
\pa_y b^{(2)}(j)=h(j)b^{(1)}(j+1)-h(j+1)b^{(1)}(j).
$$
The following Proposition is established straightforwardly by induction.
\begin{proposition}
Nonlocal variables $b^{(k)}(j)$, where $k=2,3,\dots$, satisfy the following identities:
\begin{eqnarray}
\label{uuu}
\left\lbrace
\begin{array}{lll}
\pa_y b^{(k)}(j)&=&h(j)b^{(k-1)}(j+1)-h(j+k-1)b^{(k-1)}(j)\\
\pa_x b^{(k)}(j)&=&b^{(k)}(j)\left(b(j+k)-b(j)\right)+b^{(k+1)}(j)-b^{(k+1)}(j-1)
\end{array}
\right.,
\end{eqnarray}
Compatibility of these equations for each $k$ is provided by the first of them, but for $k+1$.
\end{proposition}

The formulas~(\ref{uuu}) were obtained by A.\,B.\,Shabat~\cite{Sh95b} in order to derive the hierarchy of
symmetries for the infinite two-dimensional Toda lattice.

The function $I=I(x,y,{\mathbf u}_x,{\mathbf u}_{xx},\dots)$ is called {\it $y$-integral} of the hyperbolic system
\begin{equation}
\label{ccc}
{\mathbf u}_{xy}=F({\mathbf u},{\mathbf u}_x,{\mathbf u}_{xx},\dots),
\end{equation}
where ${\mathbf u}=(u_1,u_2,\dots,u_r)$, if its total derivative with respect to $y$ vanishes on solutions of~(\ref{ccc}): $D_y(I)=0$
(here we assume that the function $I$ depends not only on the variable $x$). $x$-integrals are defined similarly.
\begin{example}
\rm
It's easy to check that the functions
\begin{equation}
\label{intli}
I=u_{xx}-\frac{1}{2}u_x^2\quad\hbox{and}\quad J=u_{yy}-\frac{1}{2}u_y^2
\end{equation}
are $y$- and $x$-integrals of the Liouville equation
\begin{equation}
\label{eqliouv}
u_{xy}=\exp u
\end{equation}
respectively.
\end{example}

Formal integrals and symmetries of the infinite Toda lattice can be expressed in terms of nonlocal variables $b^{(k)}(j)$. For example, the formulas
\begin{eqnarray}
\nonumber
q_t(j)&=&b^2 (j)+b^{(1)}(j)+b^{(1)}(j-1),\\
\nonumber
q_t(j)&=&b^3 (j)+b^{(2)}(j)+b^{(2)}(j-1)+b^{(2)}(j-2)+\\
\nonumber
&+&b^{(1)}(j)\left(2b(j)+b(j+1)\right)+b^{(1)}(j-1)\left(2b(j)+b(j-1)\right)
\end{eqnarray}
define symmetries of the lattice~(\ref{qtoda}), and the functions
$$
I_2=\sum\limits_{j\in\mathbb Z}\left(b^2 (j)+2b^{(1)}(j)\right)\quad\hbox{and}\quad
I_3=\sum\limits_{j\in\mathbb Z}\left(b^3 (j)+3b^{(1)}(j)(b(j)+b(j+1))+3b^{(2)}(j)\right)
$$
are its $y$-integrals (see~\cite{Sh95b}). However, it's very important that all nonlocalities become local for $A$--$D$ Toda lattices. More precisely, the
following theorem holds.
\begin{theorem}
If the infinite Toda lattice is reduced by the boundary conditions~(\ref{aseries}) corresponding to Lie algebras of the series $A$, then all nonlocal
variables can be expressed in terms of the dynamic variables $b(j)$ and their multiple derivatives with respect to $x$.
\end{theorem}
{\bf Proof}.

It's easy to notice that due to the form of the equations~(\ref{uuu}) if the functions $b^{(k)} (j)$ are known for a certain $j$, then all nonlocalities
$b^{(k)}(j+1)$ can be expressed in terms of these functions and in terms of the dynamic variables. This, in turn, allows to express $b^{(k)}(j+2)$, etc. Since
the cut-off constraint~(\ref{aseries}) takes the form $h(-1)=h(r)=0$ in terms of the variables $h(j)$, the following relations hold:
$$
0=\pa_y b^{(1)}(-1)=\pa_y b^{(2)}(-1)=\pa_y b^{(3)}(-1)=\dots
$$
Therefore, if one sets
\begin{equation}
\label{ddd}
0=b^{(1)}(-1)=b^{(2)}(-1)=b^{(3)}(-1)=\dots,
\end{equation}
on the left edge, this won't lead to a contradiction. Besides this, it allows to express consequently $b^{(1)}(j)$ in terms of the dynamic variables, then
to express the functions $b^{(2)}(j)$ in terms of the dynamic variables and the nonlocalities that were already found, etc. This completes the proof.
\begin{example}
\rm
If one assumes that relations~(\ref{ddd}) are satisfied, then it's easy to verify that the following explicit formulas for nonlocalities hold for an
$A$-series lattice:
\begin{equation}
\label{pbb}
b^{(1)}(j)=\sum\limits_{i=0}^j b_x (i),\quad b^{(2)}(j)=\sum\limits_{i=0}^j \left((j-i+1)b_{xx}(i)+b_x (i) (b(i)-b(j+1))\right),
\end{equation}
where $j=0,1,\dots,r$.
\end{example}
\begin{proposition}
Generalized Toda lattices of the series $B$--$D$ are reductions of an $A$-series lattice.
\end{proposition}
{\bf Proof}.

In terms of the variable $h(j)$ the cut-off constraints~(\ref{bseries}) take the form $h(-1)=h(0)$. It's easy to verify that this implies the following
relations:
$$
q(-j)=-q(j),\quad h(-j)=h(j-1),\quad j=1,2,\dots,r+1.
$$
This involution is a reflection about the half-integer point $-1/2$ in terms of the variable $h$ and it reduces the $A$-series lattice of the length $2r+1$
defined by the condition $h(-r-1)=h(r)=0$ to the $B$-series lattice of the length $r$. Therefore a $B$-series lattice is a reduction of an $A$-series lattice.
In the same way, the cut-off constraints~(\ref{cseries}) lead to the involution
$$
q(-j)=-q(j+1),\quad h(-j)=h(j),\quad j=0,1,\dots,r,
$$
i.e. the $C$-series lattice of the length $r$ is a reduction of the $A$-series lattice of the length $2r$, defined by the constraint $h(-r)=h(r)=0$. In the case
of a $D$-series lattice the situation is more complicate because explicit formulas~(\ref{dseries}) do not imply the fact that it is a reduction of an $A$-series
lattice. For a long time this was on open problem and positive solution was given by Habibullin in~\cite{Ha05}.
\begin{corollary}
If the infinite Toda lattice is reduced by any of the boundary conditions~(\ref{bseries})--(\ref{dseries}), then all nonlocal variables can be expressed in terms of the
dynamic variables $b(j)$ and their multiple derivatives with respect to $x$.
\end{corollary}

Various approaches allowing to obtain integrals along characteristics for Toda lattices corresponding to simple Lie algebras were proposed by different
authors (see~\cite{De,GZ}). Complete families of $y$-integrals for all lattices of the series $A$--$D$ were constructed using wronskian formulas in~\cite{De}. Formulas
for $x$- and $y$-integrals were obtained in~\cite{GZ} using generalized Laplace invariants of the Liouville-type equations. However our approach based on the use of
nonlocal variables allows to obtain $y$-integrals for the lattices of the series $A$--$C$ in a very simple way and it can be extended to the semidiscrete case. Namely,
the following theorem holds.
\begin{theorem}
Functions $b^{(k)}(r)$ corresponding to any generalized lattice of the series $A$--$D$ are $y$-integrals of this lattice.
\end{theorem}
{\bf Proof}.

We have already proved that all nonlocal variables for these lattices can be expressed in terms of the dynamic variables and their derivatives with
respect to $x$. The fact that the functions $b^{(k)}(r)$ are $y$-integrals is obvious. Indeed, the equation $h(j)=0$ is satisfied for all $j\geqslant r$ for an
$A$-series lattice. Hence the relation $\pa_y b^{(k)}(r)=0$ is a direct consequence of the first formula in~(\ref{uuu}). For all other lattices the claim is an
implication of the fact that they are reductions of an $A$-series lattice.
\begin{remark}
\rm
Since the Toda lattice is symmetric about the interchange of variables $x\leftrightarrow y$ in the continuous case, $x$-integrals are obtained similarly. More
precisely, one has just to interchange variables $x$ and $y$ in the formulas for $y$-integrals.
\end{remark}
\begin{definition}
\rm
Suppose a $y$-integral $I$ of a system of hyperbolic equations of the form~(\ref{ccc}) depends on multiple derivatives of the functions $u_j$ with respect to $x$,
where $j=1,2,\dots,r$, and $d$ is the maximal order of differentiation. Then the number $d$ is called {\it the order} of the $y$-integral $I$. The family
$I_1,I_2,\dots,I_k$ of $y$-integrals of orders $d_1,d_2,\dots d_k$ respectively is called {\it essentially independent\footnote{Terminology is taken from~\cite{SS}.}}
if the rank of the matrix
\begin{eqnarray}
\label{pgg}
\frac{\pa(I_1,I_2,\dots,I_k)}{\pa ({\bf u}_{x\dots x}^{(d_1)}, {\bf u}_{x\dots x}^{(d_2)},\dots, {\bf u}_{x\dots x}^{(d_k)})}=
\left(
\begin{array}{cccc}
\frac{\pa I_1}{\pa u_{1,x\dots x}^{(d_1)}} & \frac{\pa I_1}{\pa u_{2,x\dots x}^{(d_1)}} & \dots & \frac{\pa I_1}{\pa u_{r,x\dots x}^{(d_1)}}\\
\frac{\pa I_2}{\pa u_{1,x\dots x}^{(d_2)}} & \frac{\pa I_2}{\pa u_{2,x\dots x}^{(d_2)}} & \dots & \frac{\pa I_2}{\pa u_{r,x\dots x}^{(d_2)}}\\
\vdots & &\ddots &\\
\frac{\pa I_k}{\pa u_{1,x\dots x}^{(d_k)}} & \frac{\pa I_k}{\pa u_{2,x\dots x}^{(d_k)}} & \dots & \frac{\pa I_k}{\pa u_{r,x\dots x}^{(d_k)}}\\
\end{array}
\right)
\end{eqnarray}
is equal to $k$. The system~(\ref{ccc}) is called {\it Darboux integrable}, if it admits complete families
$$
I_1,I_2,\dots,I_r\quad\hbox{and}\quad J_1,J_2,\dots,J_r
$$
of independent $y$-integrals and independent $x$-integrals respectively.
\end{definition}
\begin{remark}
\rm
Let us remark that the notion of Darboux integrability is sometimes referred to as the existence of an explicit formula for the general solution of
the equation considered (see~\cite{Ts}). This terminology is, in a sense, more reasonable since Darboux himself is more likely to have used that
very approach. Nevertheless we'll call Darboux integrability the existence of complete families of integrals in both directions. In general, these
approaches are not equivalent, but the existence of sufficient number of $x$- and $y$-integrals allows to reduce the initial partial differential
equation to a system of ordinary differential equations that can be sometimes integrated using quadratures. For instance, use of the
integrals~(\ref{intli}) allows to obtain the general solution
$$
u(x,y)=\ln\frac{2f'(x)g'(y)}{(f(x)+g(y))^2}
$$
for the Liouville equation~(\ref{eqliouv}), where $f$ and $g$ are arbitrary functions of one variable. Besides, there is another approach to integrability
of hyperbolic differential equations using the finiteness of the Laplace series for the linearized equation~\cite{Ts,ZS}.
\end{remark}

Let us give the explicit formulas for two most simple $y$-integrals of orders $1$ and $2$ respectively for the $A$-series Toda lattice of an arbitrary length:
\begin{equation}
\label{pkk}
b^{(1)}(r)=\sum\limits_{j=0}^r b_x(j),\quad b^{(2)}(r)=\sum\limits_{j=0}^{r} \left((r-j+1)b_{xx}(j)+b_x (j)b(j))\right).
\end{equation}
It is easy to verify that both integrals are total derivatives with respect to $x$. Therefore in both cases the order can be reduces by one. It can be proved
that all $y$-integrals obtained using this method are total derivatives with respect to $x$.

Darboux integrability of finite Toda lattices is widely known. Let us formulate the theorem and demonstrate that our approach allows to obtain it very simply.
\begin{theorem}
Two-dimensional Toda lattices of the series $A$, $B$ and $C$ are Darboux integrable.
\end{theorem}
{\bf Proof}.

We have already proved that the functions $b^{(k)}(r)$ are $y$-integrals for the $A$-series Toda lattice of the length $r$. Let us show that they are
independent for $k=1,2,\dots r$. Since the explicit formulas for the integrals $b^{(k)}(r)$ are rather cumbersome even for a small $k$, we won't attempt to
obtain them. We'll only examine the dependence of these integrals on the derivatives of the highest order. It is clear that the order of the integral
$b^{(k)}(r)$ equals $k$ and that this integral depends linearly on the derivatives of the highest order. Hence for all $j=0,1,\dots,r$ the nonlocalities are
expressed in terms of the dynamic variables as follows:
\begin{equation}
\label{pee}
b^{(k)}(j)=\sum\limits_{i=0}^j a^{(k)}_{ij} \frac{\pa^k b(i)}{\pa x^k}+\dots,
\end{equation}
where $a^{(k)}_{ij}$ are certain coefficients. The formula~(\ref{uuu}) implies that these coefficients are constant (i.e. that they do not depend on $x$
and on $y$) and that they satisfy the following recurrence relations:
$$
a^{(k)}_{ij}=a^{(k)}_{ij-1}+a^{(k-1)}_{ij}\quad\hbox{for}\quad i<j,\quad a^{(k)}_{jj}=a^{(k-1)}_{jj},\quad a^{(k)}_{ij}=0\quad\hbox{for}\quad i>j.
$$
In addition, it follows from the first of the formulas~(\ref{pbb}) that $a^{(1)}_{ij}=1$ for all $i\leqslant j$. This means that if we put the
coefficients $a^{(k)}_{ij}$ for a fixed $i$ into a matrix, with $k$ being a row index and $j$ being a column index, we obtain a part of the Pascal's
triangle that was rotated counterclockwise by $45^\circ$:
\begin{eqnarray}
\nonumber
i=0:\ \left(
\begin{array}{ccccc}
1 & 1 & 1 & 1 &\dots\\
1 & 2 & 3 & 4 &\dots\\
1 & 3 & 6 & 10 &\dots\\
1 & 4 & 10 & 20 &\dots\\
1 & 5 & 15 & 35 &\dots\\
\vdots & \vdots & \vdots & \vdots &\ddots
\end{array}
\right),\quad i=1:\ \left(
\begin{array}{ccccc}
0 & 1 & 1 & 1 & \dots\\
0 & 1 & 2 & 3 & \dots\\
0 & 1 & 3 & 6 & \dots\\
0 & 1 & 4 & 10 & \dots\\
0 & 1 & 5 & 15 & \dots\\
\vdots & \vdots & \vdots & \vdots &\ddots
\end{array}
\right),\quad i=2:\ \left(
\begin{array}{ccccc}
0 & 0 & 1 & 1 &  \dots\\
0 & 0 & 1 & 2 &  \dots\\
0 & 0 & 1 & 3 &  \dots\\
0 & 0 & 1 & 4 &  \dots\\
0 & 0 & 1 & 5 &  \dots\\
\vdots & \vdots & \vdots & \vdots &\ddots
\end{array}
\right),\dots
\end{eqnarray}
However we are interested only in the functions $b^{(k)}(r)$. Hence in each of these matrices we need only the last column. Therefore we have:
\begin{eqnarray}
\label{pcc}
\frac{\pa\left(b^{(1)}(r),b^{(2)}(r),\dots,b^{(r+1)}(r)\right)}{\pa\left({\bf b}_x, {\bf b}_{xx},\dots, {\bf b}_{x\dots x}^{(r+1)}\right)}=
\left(
\begin{array}{ccccc}
1 & 1 & \dots & 1 & 1\\
r+1 & r & \dots & 2 & 1\\
\frac{(r+1)(r+2)}{2} & \frac{r(r+1)}{2} & \dots & 3 &1\\
\vdots & \vdots & \vdots & \ddots &\vdots\\
\dots & \dots & \dots & r+1 & 1
\end{array}
\right).
\end{eqnarray}
It's not hard to prove that the determinant of this matrix equals $\pm 1$ depending on the parity of the number $r$. Thus the obtained family of $y$ integrals
is independent. Since the Toda lattice is symmetric about the interchange the of variables $x\leftrightarrow y$, similar construction provides the
complete family of independent $x$-integrals. Therefore $A$-series Toda lattices are Darboux integrable.

Due to the fact that Toda lattices corresponding to Lie algebras of the series $B$ and $C$ are obtained from $A$-series lattice by reduction, we only need to
show how to choose complete family of integrals so that after reduction it will remain independent. Consider a $B$-series lattice first. Since it is obtained
from the $A$-series lattice of the length $2r+1$ by the reduction
\begin{equation}
\label{pff}
q(0)=0,\quad q(-j)=-q(j),\quad h(-j)=h(j-1),\quad j=1,2,\dots,r+1,
\end{equation}
for all $j=1,2,\dots,r$ and for any $k$ we have:
\begin{equation}
\label{pdd}
\frac{\pa^k b(-j)}{\pa x^k}=-\frac{\pa^k b(j)}{\pa x^k}.
\end{equation}
The reduction we examine corresponds to the change of variables from the set $b(-r),\dots,b(r)$ to the set $b(1),b(2),\dots b(r)$. Taking into account the
formulas~(\ref{pdd}),(\ref{pee}) we can see that such change of variables affects the corresponding square matrix~(\ref{pcc}) of the size $2r+1$ in the following way:
the column with the number $r-j$ is subtracted from the column with the number $r+j$ for all $j=1,2,\dots,r$. Since the columns of the initial matrix are
independent the columns of the new $(2r+1)\times r$-matrix are also independent. This means that it's possible to choose $r$ rows in this matrix so that the
corresponding minor will be non-zero. Thus the $B$-series lattice obtained using the reduction~(\ref{pff}) admits a complete family of independent
$y$-integrals. Consequently due to the symmetry between $x$ and $y$ this implies that $B$-series lattices are Darboux integrable.

Consider now a $C$-series lattice. It is obtained using the reduction
$$
q(-j)=-q(j+1),\quad h(-j)=h(j),\quad j=0,1,\dots,r.
$$
from the $A$-series lattice of the length $2r$. In the same way, this reduction corresponds to the change of variables from the set $b(-r+1),\dots,b(r)$
to the set $b(1),b(2),\dots b(r)$, and this yields that the column with the number $r-j-1$ is subtracted from the column with the number $r+j$ for all
$j=0,1,\dots,r-1$ in the corresponding $2r\times 2r$-matrix. By the same argument, this implies Darboux integrability of $C$-series lattices. This concludes
the proof.
\begin{remark}
\rm
Since there are no explicit formulas defining a reduction of an $A$-series lattice to a $D$-series lattice, similar approach can not be applied to the $D$-series
case due to the fact that integrals can possibly become dependent under such reduction.
\end{remark}

\subsection{Generating function for integrals}

Let us describe now another construction allowing to obtain $x$- and $y$-integrals for finite Toda lattices. The peculiarity of this construction is that the
existence of integrals is a direct consequence of the fact that Toda lattice is the relation on Laplace invariants of hyperbolic operators in a sequence where
any two neighboring terms are connected by Darboux--Laplace transformations. Trivial boundary conditions $h(-1)=h(r)=0$ for an $A$-series lattice provide an
opportunity to close this sequence at both edges with operators that can be factorized. As an example, very similar construction allows to obtain first integrals for
periodic closure of the Veselov--Shabat dressing chain that describes a sequence of one-dimensional Schr\"odinger operators connected by Darboux
transformations~\cite{VS}. In this case, the generating function for first integrals is also obtained from the operator relation provided by Darboux
transformations.
\begin{theorem}
Coefficients of the differential operator
\begin{equation}
\label{pll}
{\mathcal B}=(\pa_x-q_x(r))(\pa_x-q_x(r-1))\dots (\pa_x-q_x(0))
\end{equation}
form a complete family of independent $y$-integrals for the $A$-series Toda lattice defined by the boundary conditions $h(-1)=h(r)=0$.
\end{theorem}
{\bf Proof}.

It can easily be checked that if the hyperbolic operators
$$
{\mathcal L}=\pa_x\pa_y+a\pa_x+b\pa_y+c\quad\hbox{and}\quad\hat{\mathcal L}=\pa_x\pa_y+\hat a\pa_x+b\hat\pa_y+\hat c
$$
are related by a Darboux--Laplace transformation of the form ${\mathcal D}=\pa_x+\gamma$, then $\hat a=a$ and $\gamma=b$ (see~(\ref{eee})). This means that in the
case we consider all operators in the sequence have the same coefficient of $\pa_x$. Therefore one can assume without loss of generality that it is equal to zero.
Besides, the equations~(\ref{eee}) imply the relation
$$
\hat b=b-(\ln h)'_x
$$
that immediately yields the condition $b(j)=-q_x(j)$ modulo addition of an arbitrary function $\phi(x,y)$ to all variables $b(j)$\footnote{Note that there is a
difference in notation here with the Section~\ref{contnonloc} where the symbol $b(j)$ stands for $q_x(j)$. However this inconvenience is inevitable because
attempts to make that notation self-consistent within this paper would cause differences in notation with one of the papers~\cite{Sh95b} or~\cite{Sm}. This
would have been more disgusting because in this case the formulas here and in one of the papers~\cite{Sh95b} or~\cite{Sm} would differ essentially.} Thus the
operator~(\ref{pll}) can be rewritten in terms of coefficients of the hyperbolic operators as follows:
$$
{\mathcal B}=(\pa_x+b(r))(\pa_x+b(r-1))\dots (\pa_x+b(0)).
$$

Since the equation $k(j)=h(j-1)$ is satisfied for all $j$, the following holds: $k(0)=h(-1)=0$. Hence ${\mathcal L}_0=(\pa_x+b(0))\pa_y$. Besides, the relation
$h(r)=0$ yields the existence of the factorization ${\mathcal L}_r=\pa_y (\pa_x+b(r))$. Therefore consequtive application of Darboux--Laplace
transformations leads to the following:
\begin{multline}
\nonumber
{\mathcal B}\circ\pa_y=(\pa_x +b(r))\dots (\pa_x+b(1))(\pa_x+b(0))\pa_y=(\pa_x +b(r))\dots (\pa_x+b(2)){\mathcal D}_1 {\mathcal L}_0=\\
\nonumber
=(\pa_x +b(r))\dots (\pa_x+b(2)){\mathcal L}_1{\mathcal D}_0={\mathcal D}_r\dots {\mathcal D}_2{\mathcal L}_1{\mathcal D}_0=\dots
={\mathcal L}_r {\mathcal D}_{r-1}{\mathcal D}_{r-2}\dots {\mathcal D}_0=\\
\nonumber
=\pa_y (\pa_x+b(r))\dots (\pa_x+b(0))=\pa_y\circ {\mathcal B}.
\end{multline}
Thus the operators ${\mathcal B}$ and $\pa_x$ commute. This implies that all coefficients of the operator ${\mathcal B}$ do not depend on $y$ if the functions
$q(j)$ are solutions of the $A$-series Toda lattice, i.e. they are $y$-integrals.

Let $I_0,I_1,\dots,I_r$ be the coefficients of the operator ${\mathcal B}$: ${\mathcal B}=\pa^{r+1}_x+I_0\pa^r_x+\dots+I_{r-1}\pa_x+I_r$. Independence of these
$y$-integrals follows from the fact that the corresponding matrix
$$
\frac{\pa\left(I_0,I_1,\dots,I_r\right)}{\pa\left({\bf b}, {\bf b}_x,\dots, {\bf b}_{x\dots x}^{(r)}\right)}
$$
is upper-triangular with respect to its secondary diagonal and all elements of this diagonal are non-zero. This completes the proof.
\begin{example}
\rm
Let us give explicitly the complete family of $y$-integrals provided by this approach for the case $r=2$. Using the formula~(\ref{pll}) we have:
\begin{align}
I_0&=-(q_x(0)+q_x(1)+q_x(2)),
\notag\\
I_1&= q_x(0)q_x(1)+q_x(0)q_x(2)+q_x(1)q_x(2)-2q_{xx}(0)-q_{xx}(1),
\notag\\
I_2&=-q_x(0)q_x(1)q_x(2)+q_{xx}(0)(q_x(1)+q_x(2))+q_{xx}(1)q_x(0)-q_{xxx}(0).
\notag
\end{align}
\end{example}
\begin{remark}
\rm
Clearly, these $y$-integrals differ from the ones obtained using the nonlocal variables in the previous Section.
\end{remark}

\section{Semidiscrete case}\label{semidisc}
\subsection{Darboux--Laplace transformations}

In the semidiscrete case as well as in the continuous one, the Laplace invariants of second order hyperbolic differential operators related by Darboux--Laplace
transformations satisfy the system of differential-difference equations called {\it the semidiscrete Toda lattice}. We'll follow the article~\cite{AS} and
use this idea to obtain the semidiscrete Toda equations.

Consider a sequence of hyperbolic differential-difference operators
$$
{\cal L}_j=\pa_x T+a_n(j)\pa_x+b_n(j) T+c_n(j),
$$
where $a_n(j)$, $b_n(j)$ and $c_n(j)$ are functions depending on discrete variable $n\in\mathbb Z$ and on continuous variable $x\in\mathbb R$
and where $T$ is a shift operator: $T\psi_n (x)=\psi_{n+1}(x)$. Obviously, the operator ${\cal L}_j$ can be factorized in two different ways:
$$
{\cal L}_j=(\pa_x+b_n(j))(T+a_n(j))+a_n(j)k_n(j)=(T+a_n(j))(\pa_x+b_{n-1}(j))+a_n(j)h_n(j),
$$
where $k_n(j)=\frac{c_n(j)}{a_n(j)}-(\ln a_n(j))'_x-b_n(j)$ and
$h_n(j)=\frac{c_n(j)}{a_n(j)}-b_{n-1}(j)$ are {\it the Laplace invariants}
of differential-difference operator ${\cal L}_j$.

Suppose any two neighboring operators ${\cal L}_j$ and ${\cal L}_{j+1}$ are related by {\it a Darboux--Laplace transformation}, that is, satisfy
the relation
$$
{\cal L}_{j+1}{\cal D}_j={\cal D}_{j+1}{\cal L}_j,
$$
where ${\cal D}_j=\pa_x+b_{n-1}(j)$. This operator relation can be rewritten in terms of coefficients as follows:
\begin{eqnarray}
\nonumber
\left\lbrace
\begin{array}{l}
k_n (j+1)=h_n(j)\\
\left(\ln\frac{h_n(j)}{h_{n+1}(j)}\right)'_x=h_{n+1}(j+1)-h_{n+1}(j)-h_n(j)+h_n(j-1)
\end{array}
\right..
\end{eqnarray}
Using the new set of variables $q_n(j)$ that is introduced by the equations $h_n(j)=\exp(q_{n+1}(j+1)-q_n(j))$, one obtains {\it the semidiscrete Toda lattice}:
\begin{equation}
\label{todaq}
q_{n,x}(j)-q_{n+1,x}(j)=\exp(q_{n+1}(j+1)-q_n (j))-\exp(q_{n+1}(j)-q_n(j-1)).
\end{equation}

Similarly to the continuous case let us introduce one more set of variables $u_n(j)$ using the relation $h_n (j)=u_{n,x}(j)-u_{n+1,x}(j)$. Then the Toda
lattice can be rewritten as follows:
\begin{equation}
\label{sdutoda}
u_{n,x}(j)-u_{n+1,x}(j)=\exp(u_{n+1}(j+1)-u_{n+1}(j)-u_n(j)+u_n (j-1)).
\end{equation}

The following semidiscrete analog of exponential systems was introduced in the paper~\cite{HZY}. Let $K=(a_{jk})$ be a square matrix. Then the system
\begin{equation}
\label{dexpsyst}
u_{n,x} (j)-u_{n+1,x} (j)=\exp\left(\sum\limits_{k=1}^{j-1} a_{jk} u_n(k)+\frac{a_{jj}}{2}(u_n (j)+u_{n+1} (j))+
\sum\limits_{k=j+1}^{r} a_{jk} u_{n+1}(k)\right),
\end{equation}
where $j=1,2,\dots,r$, is a natural semidiscrete analog of the exponential system~(\ref{expsyst}) corresponding to the matrix $K$. Clearly, the expression
inside the exponent in the formula~(\ref{dexpsyst}) corresponds to the presentation of the matrix $K$ as a sum of lower-triangular and upper-triangular
matrices such that diagonal elements in these matrices are the same. Semidiscrete exponential system~(\ref{dexpsyst}) tends to its continuous analog in the
continuum limit.

In the semidiscrete case as well as in the continuous one it's natural to examine the reductions of the infinite Toda lattice. It is easily shown that the
trivial boundary conditions $h_n(-1)=h_n(r)=0$ lead to exponential systems corresponding to the Cartan matrices of $A$-series Lie algebras. In terms of the variable
$q$ such closure is defined by the cut-off constrains $q_n(-1)=\infty$, $q_n (r+1)=-\infty$. And in terms of the variable $u$ the series $A$ corresponds to
the system~(\ref{sdutoda}) with trivial boundary conditions $u(0)=u(r+1)=0$.

Let us examine what involutions does the Toda lattice admit. Consider the equations in terms of the Laplace invariants, i.e. the system
\begin{equation}
\label{sdhtoda}
\left(\ln\frac{h_n(j)}{h_{n+1}(j)}\right)'_x=h_{n+1}(j+1)-h_{n+1}(j)-h_n(j)+h_n(j-1).
\end{equation}
It can easily be checked that the reflection $h_n(-j)=h_{n+j+c}(j-d)$ defines a reduction of the lattice~(\ref{sdhtoda}) only if $d=-2c$. This means
that the situation here differs from the one in the continuous case, where both reflections about half-integer and about integer points generate the
reductions of the infinite lattice corresponding to $B$-series and $C$-series respectively. The choice $c=0$ leads to the boundary condition $h_n(-j)=h_{n+j}(j)$
(this equation should be satisfied for all $n\in\mathbb Z$). If $j=1$, then this gives the following cut-off constraint in terms of the variable $q$:
\begin{equation}
\label{todac}
q_{n+1} (0)-q_n(-1)=q_{n+2}(2)-q_{n+1}(1).
\end{equation}
The lattice~(\ref{todaq}) satisfying the cut-off constraint~(\ref{todac}) on the left edge and the trivial boundary condition
$q_n(r+1)=-\infty$ on the right edge leads to exponential systems corresponding to the $C$-series Lie algebras. Exactly as it is in the continuous case,
$C$-series semidiscrete lattice is a reduction not only of the infinite lattice but it's also a reduction of the $A$-series lattice of the double length. Indeed, it
is easily shown that the involution $h_n(-j)=h_{n+j}(j)$ reduces the $A$-series lattice defined by the constraints $h_n (-r)=h(r)=0$ to the $C$-series lattice.
The question of whether the semidiscrete lattices~(\ref{dexpsyst}) corresponding to Lie algebras of $B$- and $D$-series are reductions of an $A$-series lattice remains
an open problem.
\begin{remark}
\rm
The relations $h_n(-j)=h_{n+j}(j)$ defining the reduction of $A$-series lattices to $C$-series lattices allow to express the variables
$q_n(-r),q_n(-r+1),\dots,q_n(-1)$ in terms of the variables
$$
q_n(0),q_n(1),\dots,q_n(r)
$$
and their shifts in $n$. However this set of variables is, in fact, not independent because the relation
$$
q_{n,x}(0)-q_{n+1,x}(0)=\exp(q_{n+1}(1)-q_n(0))-\exp(q_{n+2}(2)-q_n(1))
$$
allows one to express $q_n(0)$ in terms of the variables $q_n(1),q_n(2),\dots,q_n(r)$ modulo constants of integration $\varkappa_n$. Without loss of generality it
can be assumed that they are zero due to the fact that the set of solutions of the Toda lattice is invariant under the substitution
$q_n(j)\rightarrow q_n(j)+\varkappa_{n-j}$.
\end{remark}

\subsection{Nonlocal variables and construction of $n$-integrals}

Symmetries of two-dimensional Toda lattices in the semidiscrete case as well as in the continuous case are
expressed in terms of nonlocal variables. Following the paper~\cite{Sm} we'll state the basic propositions and formulas for the semidiscrete
case that are similar to the ones from the Subsection~\ref{contnonloc}. Introduce the notation $\pa_n=I-T$,
where $I$ is the identity operator and $b_n (j)=q_{n,x}(j)$. Then the lattice~(\ref{todaq}) can be rewritten as
follows:
\begin{equation}
\label{aaa}
\pa_n b_n (j)=h_n(j)-h_n(j-1).
\end{equation}
Define the nonlocal variables $b_n^{(1)}(j)$ by the following formulas:
$$
\pa_x b_n (j)=b_n^{(1)}(j)-b_n^{(1)}(j-1),\quad\pa_n b_n^{(1)} (j)=\pa_x h_n(j).
$$
Compatibility of these equations is provided by the Toda lattice~(\ref{aaa}):
$$
\pa_n\pa_x b_n (j)=\pa_n \left(b_n^{(1)}(j)-b_n^{(1)}(j-1)\right)=\pa_x \left(h_n (j)-h_n (j-1)\right)=\pa_x\pa_n b_n (j).
$$
The next step is to determine the derivatives $\pa_x b_n^{(1)}(j)$ but this, however, requires to introduce nonlocalities $b_n^{(2)} (j)$
of the second order. Let
$$
\pa_x b_n^{(1)} (j)=b_n^{(1)}(j)\left(b_{n+1}(j+1)-b_n(j)\right)+h_n(j)\left(b_n^{(1)}(j-1)-b_n^{(1)}(j)\right)+b_n^{(2)}(j)-b_n^{(2)}(j-1);
$$
then the compatibility condition $\pa_n\pa_x b_n^{(1)}(j)=\pa_x\pa_n b_n^{(1)}(j)$ obviously leads to the following relation:
$$
\pa_n b_n^{(2)}(j)=h_n(j)b_{n+1}^{(1)}(j+1)-h_{n+1}(j+1)b_{n+1}^{(1)}(j).
$$
The following Proposition is proved by standard inductive reasoning.
\begin{proposition}
Nonlocal variables $b_n^{(k)}(j)$, where $k=2,3,\dots$, satisfy the following equations:
\begin{eqnarray}
\label{bbb}
\left\lbrace
\begin{array}{lll}
\pa_n b_n^{(k)}(j)&=&h_n (j)b_{n+1}^{(k-1)}(j+1)-h_{n+k-1} (j+k-1)b_{n+1}^{(k-1)}(j)\\
\pa_x b_n^{(k)}(j)&=&b_n^{(k)}(j)\left(b_{n+k}(j+k)-b_n(j)\right)+\\
&+&h_{n+k-1} (j+k-1)\left(b_n^{(k)}(j-1)-b_n^{(k)}(j)\right)+b_n^{(k+1)}(j)-b_n^{(k+1)}(j-1)
\end{array}
\right.,
\end{eqnarray}
The compatibility of these equations for each $k$ is provided by the first of them, but for $k+1$.
\end{proposition}

We won't consider symmetries of the infinite semidiscrete Toda lattice and of its reductions in this paper (see~\cite{Sm}). Here we'll concentrate on
$n$-integrals and $x$-integrals of this system. The function $I$ depending on dynamic variables is called {\it an $n$-integral} of a differential-difference system
if its difference derivative vanishes: $\pa_n I=0$. We are going to demonstrate how to obtain a complete family of $n$-integrals for semidiscrete lattices
corresponding to $A$-series and to $C$-series using nonlocal variables. In the next Section we'll give an explicit formula for $x$-integrals of these systems.
\begin{definition}
\rm
Suppose an $n$-integral $I$ of a system of differential-difference equations depends on multiple derivatives of the functions $u_n(j)$ with respect to $x$,
where $j=1,2,\dots,r$, and $d$ is the maximal order of differentiation. Then the number $d$ is called {\it the order} of a $n$-integral $I$. The family
$I_1,I_2,\dots,I_k$ of $y$-integrals of orders $d_1,d_2,\dots d_k$ respectively is called {\it essentially independent}
if the rank of the corresponding matrix~(\ref{pgg}) is equal to $k$. Suppose an $x$-integral $J$ of a system of differential-difference equations depends on
the shifts $u_{n+d} (j)$ of dynamical variables $u_n(j)$, where $j=1,2,\dots,r$, and $d$ is the maximal shift (here we assume that this integral depends on
$u_n (j)$ and that it does not depend on $u_{n-1}(j),u_{n-2}(j),\dots$). Then the number $d$ is called {\it the order} of the $x$-integral $J$. The family
$J_1,J_2,\dots,J_k$ of $x$-integrals of orders $d_1,d_2,\dots d_k$ respectively is called {\it essentially independent}
if the rank of the matrix
\begin{eqnarray}
\label{ptt}
\frac{\pa(I_1,I_2,\dots,I_k)}{\pa ({\bf u}_{n+d_1}, {\bf u}_{n+d_2},\dots, {\bf u}_{n+d_k})}=
\left(
\begin{array}{cccc}
\frac{\pa I_1}{\pa u_{n+d_1} (1)} & \frac{\pa I_1}{\pa u_{n+d_1}(2)} & \dots & \frac{\pa I_1}{\pa u_{n+d_1}(r)}\\
\frac{\pa I_2}{\pa u_{n+d_2} (1)} & \frac{\pa I_2}{\pa u_{n+d_2}(2)} & \dots & \frac{\pa I_2}{\pa u_{n+d_2}(r)}\\
\vdots & &\ddots &\\
\frac{\pa I_k}{\pa u_{n+d_k} (1)} & \frac{\pa I_k}{\pa u_{n+d_k}(2)} & \dots & \frac{\pa I_k}{\pa u_{n+d_k}(r)}\\
\end{array}
\right)
\end{eqnarray}
equals $k$. Differential-difference system is called {\it Darboux integrable} if it admits complete families
$$
I_1,I_2,\dots,I_r\quad\hbox{and}\quad J_1,J_2,\dots,J_r
$$
of independent $n$-integrals and independent $x$-integrals respectively.
\end{definition}

The following two theorems are proved similarly to the continuous case.
\begin{theorem}\label{thint}
If the infinite Toda semidiscrete lattice~(\ref{todaq}) is reduced by imposing of trivial boundary conditions corresponding to the series $A$, then all nonlocal
variables can be expressed in terms of the dynamic variables $b_n(j)$ and their multiple derivatives with respect to $x$.
\end{theorem}
The next claim follows from the fact that the semidiscrete lattice of the length $r$ corresponding to the series $C$ is a reduction of the $A$-series lattice
of the length $2r+1$.
\begin{corollary}
In the case of the series $C$ all nonlocal variables can be expressed in terms of the dynamic variables $b_n(j)$ and their multiple derivatives with
respect to $x$.
\end{corollary}
\begin{theorem}
Functions $b_n^{(k)}(r)$ corresponding to any generalized lattice of the series $A$ and $C$ are $n$-integrals of this lattice.
\end{theorem}

Let us give the explicit formulas for two most simple $n$-integrals of an $A$-series lattice of an arbitrary length $r$:
\begin{equation}
\label{phh}
b_n^{(1)}(r)=\sum\limits_{j=0}^r b_{n,x}(j),\quad b^{(2)}_n(r)=\sum\limits_{j=0}^{r} \left((r-j+1)b_{n,xx}(j)+b_{n,x} (j)b_n(j))\right).
\end{equation}
\begin{remark}
\rm
Obviously, the explicit formulas~(\ref{phh}) for $n$-integrals of $A$-series lattices are the same as the formulas~(\ref{pkk}) for $y$-integrals in the
continuous case, and this is not occasional. Usually when discrete analogs of well-known integrable systems are constructed, discretizations having the same
integrability attributes such as integrals or symmetries are considered. Exactly this empiric principle was used in the paper~\cite{HZY} for constructing
semidiscrete exponential systems. The authors were looking for semidiscrete systems (for small $r$) such that $y$-integrals of the corresponding continuous
analogs are $n$-integrals of these systems.
\end{remark}
\begin{remark}
\rm
The notion of characteristic algebra for the semidiscrete exponential systems is introduced in the papers~\cite{HZY,HP}. Similarly to the continuous case,
finite dimensionality of the characteristic algebra in one of the directions is equivalent to the existence of integrals in this direction. Characteristic
algebras of semidiscrete exponential systems for certain small $r$ are described explicitly in the paper~\cite{HZY}. In particular, it follows from the
Theorem~\ref{thint} that the characteristic algebra in the direction $n$ is finite dimensional for any of the lattices corresponding to Lie algebras of the
series $A$ and $C$.
\end{remark}
\begin{remark}
\rm
It's rather natural to expect that there is another set of nonlocal variables for the semidiscrete Toda lattice that is, in a sense, a mirror-image of the set
$b_n^{(k)}(j)$ about the interchange of the discrete variable $n$ and continuous variable $x$ and that such set of nonlocal variables would have allowed to
obtain $x$-integrals for $A$-series lattices. However in fact everything is not so simple. If such construction is possible, it is not obvious. Constructions
parallel to the ones above meet obstacles when one attempts to obtain the second set of nonlocalities.
\end{remark}

\subsection{Explicit formulas for integrals}

In the semidiscrete case it's also possible to obtain generating functions for $n$-integrals and for $x$-integrals of $A$-series lattices and their reductions
exactly in the same way as in the continuous case, and the existence of such generating function is a consequence of the structure of the Toda lattice associated
with Darboux--Laplace transformations. However one has to consider Darboux--Laplace transformations of two different types since the variables $n$ and $x$ enter the
equations of the semidiscrete lattice not symmetrically. It can easily be checked that the operator relation
$\hat{\mathcal L}{\mathcal D}=\hat{\mathcal D}{\mathcal L}$ where ${\mathcal D}=\alpha_n\pa_x+\beta_n T+\gamma_n$ and
$\hat{\mathcal D}=\hat\alpha_n\pa_x+\hat\beta_n T+\hat\gamma_n$ is equivalent to the following system of differential-difference equations:
\begin{eqnarray}
\label{fff}
\left\lbrace
\begin{array}{l}
\alpha_{n+1}-\hat\alpha_n=0\\
\beta_{n+1}-\hat\beta_n=0\\
\alpha_n\hat a_n-\hat\alpha_n a_n=0\\
\beta_{n+1,x}+\beta_{n+1}\hat b_n-\hat\beta_n b_{n+1}=0\\
\alpha_{n+1,x}+\alpha_{n+1}\hat b_n-\hat\alpha_n b_n+\beta_n\hat a_n-\hat\beta_n a_{n+1}+\gamma_{n+1}-\hat\gamma_n=0\\
\hat a_n \alpha_{n,x}+\hat c_n\alpha_n-\hat\alpha(c_n+a_{n,x})+\hat a_n\gamma_n-\hat\gamma_n a_n=0\\
\hat a_n \beta_{n,x}+\hat c_n\beta_n-\hat\beta_n c_n-\hat\alpha_n b_{n,x}+\gamma_{n+1,x}+\hat b_n\gamma_{n+1}-\hat\gamma_n b_n=0\\
\hat a_n \gamma_{n,x}+\hat c_n\gamma_n-\hat\alpha_n c_{n,x}-\hat\gamma_n c_n=0
\end{array}
\right..
\end{eqnarray}
The choice of parameters $\alpha_n\equiv 1,\beta_n\equiv 0$ corresponding to the Darboux--Laplace transformations of the first type (with respect to the
continuous variable) yields the relation $\gamma_n=b_{n-1}$. We've already mentioned that for a sequence of Darboux--Laplace transformations the semidiscrete
Toda lattice equations~(\ref{sdhtoda}) for the variable $h$ being a Laplace invariant follow from such system. It appears that the equations $k_n (j+1)=h_n(j)$
and $a_n(j+1)=a_n(j)$ are satisfied for all $j$. The second of these equations allows to assume that all coefficients $a_n(j)$ are equal to some constant
$\varkappa_n$\footnote{We can not assume these coefficients to be zeroes as we did in the continuous case because there is division by $a_n$ in the definition
of the Laplace invariants in the semidiscrete case.}. By the same argument we claim that the operators
${\mathcal B}={\mathcal D}_r{\mathcal D}_{r-1}\dots{\mathcal D}_0$ and $T+\varkappa_n I$, where $I$ is the identity operator, commute. This, in turn, implies
that the operators ${\mathcal B}$ and $T$ also commute. Besides, the equations~(\ref{fff}) yield that
$$
\hat b_n=b_{n-1}-\left(\ln h_n\right)'_x.
$$
Now it follows easily that $b_n (j)=-q_{n+1,x}(j)$. For the same reason as in the continuous case this leads to the following theorem.
\begin{theorem}
The coefficients of the differential operator
$$
{\mathcal B}=(\pa_x -q_{n,x}(r))(\pa_x-q_{n,x}(r-1))\dots (\pa_x-q_{n,x}(0))
$$
form a complete family of independent $n$-integrals for the semidiscrete Toda lattice of the series $A$ defined by the constraints $h_n(-1)=h_n(r)=0$.
\end{theorem}

Consider now Darboux--Laplace transformations of the other type corresponding to the choice of the parameters $\alpha_n\equiv 0,\beta_n\equiv 1$. In this case the
system~(\ref{fff}) yields the relation $\gamma_n=a_n$. Besides, without loss of generality one may assume the coefficients $b_n$ to be zeroes, as before. The
difference with the above case is that the corresponding system of differential-difference equation in this case yields the following equation for the Laplace
invariants:
\begin{equation}
\label{sdktoda}
\left(\ln\frac{k_n(j)}{k_{n+1}(j)}\right)'_x=k_{n+1}(j-1)-k_{n+1}(j)-k_n(j)+k_n(j+1).
\end{equation}
Likewise, $h_n (j+1)=k_n(j)$ for all $j$. It follows easily that the system~(\ref{sdktoda}) is equivalent to the Toda lattice~(\ref{sdhtoda}) after the
substitution $j\leftrightarrow -j$. Hence Darboux--Laplace transformations of the second type also lead to the semidiscrete Toda lattice. Let us introduce the
new set of variables $p_n (j)$ using the relation
$$
k_n(j)=\exp(p_{n+1}(j)-p_n (j+1)).
$$
The system~(\ref{sdktoda}) can be rewritten as follows:
$$
p_{n,x}(j)-p_{n+1,x}(j)=k_n (j-1)-k_n (j).
$$
It is clear that this equation turns into~(\ref{todaq}) after the substitution $q_n(j)=p_n(-j)$. Note that in this case the system~(\ref{fff}) implies the
following equation:
$$
\hat a_n=\frac{a_{n+1}k_{n+1}}{k_n}.
$$
Now it follows easily that $a_n(j)=\exp(p_n(j)-p_{n+1}(j))$. Combining all these arguments together we can see that Darboux--Laplace transformations allow to
obtain a generating function for $x$-integrals also. The only difference is that the order of factors in the corresponding operator will be different in this
case due to the involution $j\leftrightarrow -j$. This proves the following theorem.
\begin{theorem}
The coefficients of the operator
\begin{equation}
\label{pmm}
{\mathcal A}=(T +\exp(q_n(0)-q_{n+1}(0)))(T+\exp(q_n(1)-q_{n+1}(1)))\dots (T+\exp(q_n(r)-q_{n+1}(r)))
\end{equation}
form a complete family of independent $x$-integrals for the $A$-series semidiscrete Toda lattice defined be the constraints $h_n(-1)=h_n(r)=0$.
\end{theorem}
\begin{remark}
\rm
It is not difficult to express the integrals provided by~(\ref{pmm}) with an explicit formula. Indeed, if we denote ${\mathcal A}=T^{r+1}+J_r T^r+\dots+J_1 T+J_0$,
then the coefficients $J_k$ are defined as follows:
\begin{multline}
\nonumber
J_k=\sum\limits_{0\leqslant i_1<i_2<\dots<i_k\leqslant r}^{}\exp (c_n(0)+\dots+c_n (i_1-1)+c_{n+1}(i_1+1)+\dots+c_{n+1}(i_2-1)+\\
+c_{n+2}(i_2+1)+\dots+c_{n+k-1}(i_r-1)+c_{n+k}(i_k+1)+\dots+c_{n+k}(r)),
\end{multline}
where $c_n(j)=q_n(j)-q_{n+1}(j)$. The most simple of these integrals (for the case $r=2$) were obtained in~\cite{HZY}.
\end{remark}
\begin{proposition}
Semidiscrete Toda lattices corresponding to Lie algebras of the series $C$ are Darboux integrable.
\end{proposition}
{\bf Proof}.

Since the $C$-series lattice is a reduction of the $A$-series lattice of the double length it is sufficient to prove that the complete families of $n$-integrals
and $x$-integrals remain independent after this reduction (more precisely, we need to prove that it is possible to choose such sub-families that are
independent for the reduced lattice). Consider the complete family of $n$-integrals for the $A$-series lattice and the corresponding $2r\times 2r$-matrix consisting
of partial derivatives first. We need to understand what happens to its $r$ right columns after the reduction. It is clear that the derivatives of the highest
order enter the expressions for $n$-integrals linearly and that corresponding coefficients are constant. Besides, the formulas that allow to express the variables
$q_n(-r),q_n(-r+1),\dots,q_n(-1)$ and their shifts with respect to $n$ in terms of the dynamic variables $q_n(1),q_n(2),\dots, q_n(r)$ and their shifts, are linear.
Nonlinearity caused by the necessity to express the derivatives of the shifted variables using the system of equations appears as a result of differentiation of these
formulas. But on the level of highest derivatives all formulas will still remain linear due to the fact that the Toda lattice equations reduce the order of
differentiation by one. This means that to each of $r$ most right columns of our matrix a linear combination of other columns will be added after the reduction.
Let us describe this linear transformation explicitly.

It is clear that for all $j=1,2\dots,r-1$ the closure constraints imply the following relations
\begin{multline}
\label{poo}
b_n(-j)=b_{n+1}(-j+1)+b_{n+j}(j)-b_{n+j+1}(j+1)=b_n(-j+1)+b_n(j)-b_n(j+1)+\dots=\\
=\dots=b_n(0)+b_n(1)-b_n(j+1)+\dots,
\end{multline}
where the dots denote the dependence on the derivatives of the functions $q_n(j)$ of zero order. Further, using the obvious relation
\begin{equation}
\label{pnn}
\pa_n\left(\sum\limits_{j=-r+1}^r b_n(j)\right)=0,
\end{equation}
one obtains that the sum in the braces in~(\ref{pnn}) does not depend on $n$, i.e. it is a function of $x$. Due to the fact that addition of a same
function $\phi(x)$ to all dynamic variables maps solutions of the Toda lattice to solutions, without loss of generality one can assume that
$$
\sum\limits_{j=-r+1}^r b_n(j)=0.
$$
Substitution of the expressions~(\ref{poo}) into this equation leads to the following relation on the level of highest derivatives:
$$
0=(r-1)(b_n(0)+b_n(1))-(b_n(2)+b_n(3)+\dots+b_n(r))+\sum\limits_{j=0}^r b_n(j)+\dots
$$
Therefore $b_n(0)=-b_n(1)+\dots$. Finally, substitution of this expression into~(\ref{poo}) yields the equation $b_n(-j)=-b_n(j+1)+\dots$. This means that the
linear transformation acting on columns of the matrix we are interested in is just the subtraction of the column with the number $j$ of the column with the number
$r+j$ for all $j=1,2,\dots,r$. Hence the $r$ columns we obtain are linearly independent and therefore one can choose a non-zero minor of the size $r$ out of
this $2r\times r$-matrix. Thus it's possible to single out $r$ independent $n$-integrals of the whole family after the reduction.

Now let us demonstrate that the family of $x$-integrals that was obtained above also remains independent after reduction. Obviously, the order of $x$-integral
$J_k$ is equal to $k$, where $k=0,1,\dots 2r$. Also, it can easily be checked that the maximal shift $d$ such that the explicit formula for $J_k$ depends
on $c_{n+d}(j)$ is defined as follows:
\begin{eqnarray}
\nonumber
d=\left\{
\begin{array}{l}
j+r-1,\quad\hbox{если}\quad j< k-r+1\\
k,\quad\hbox{если}\quad j\geqslant k-r+1
\end{array}
\right.,
\end{eqnarray}
where $j=-r+1,\dots,r$. Let us prove that the integrals $J_{r+1},J_{r+2},\dots J_{2r}$ remain independent after the reduction to the $C$-series lattice. Indeed,
the closure condition~(\ref{todac}) yields the following relation:
$$
c_n(-j)=\sum\limits_{i=0}^j c_{n+j}(i)-\sum\limits_{i=2}^{j+1} c_{n+j+1} (i),
$$
where $j=1,2,\dots,r-1$. This means that after the reduction the variables $c_{n+d}(j)$ with the maximal shift $d$ will appear in the expressions for integrals
$J_k$ where $d=n+r$ if $j=2,3,\dots r$, and $d=n+r-1$ if $j=0,1$. But the variable $c_n(0)$ can be expressed using the relation
$$
c_{n,x}(0)=h_n(0)-h_{n+1}(1),
$$
and this substitution does not increase the order of integrals. Therefore the reduction does not affect the highest shifts in the expressions for $x$-integrals
$J_{r+1},J_{r+2},\dots J_{2r}$. Since the matrix~(\ref{ptt}) is upper-triangular for the $A$-series lattice, the corresponding matrix after reduction is also
upper-triangular and hence nonsingular. Thus the $x$-integrals $J_{r+1},J_{r+2},\dots J_{2r}$ appear to be independent after reduction to the series $C$.
This concludes the proof.

\section{Purely discrete case}\label{disc}

Similarly to the semidiscrete case in the purely discrete case the Laplace invariants for a hyperbolic second order differential operator are also defined, and
the sequence of such operators connected by Darboux--Laplace transformations leads to the purely discrete two-dimensional Toda lattice. More precisely,
the system of equations for the coefficients of these operators after elimination of some variables is equivalent to difference equations on the Laplace
invariants that are called {\it the purely discrete Toda lattice}. Let us give the main formulas following the paper~\cite{AS}.

Consider a sequence of hyperbolic difference operators
$$
{\cal L}_j=T_1 T_2+a(j)T_1+b(j)T_2+c(j),
$$
where $a(j)$, $b(j)$ and $c(j)$ are functions depending on two discrete variables $n,m\in\mathbb Z$ and $T_1$, $T_2$ are the shift operators:
$$
T_1\psi_{n,m}=\psi_{n+1,m}\quad T_2\psi_{n,m}=\psi_{n,m+1}.
$$
It is clear that the operator ${\cal L}_j$ can be factorized in two different ways:
\begin{multline}
\nonumber
{\cal L}_j=(T_1+b_{n,m}(j))(T_2+a_{n-1,m}(j))+a_{n-1,m}(j)b_{n,m}(j)k_{n,m}(j)=\\
=(T_2+a_{n,m}(j))(T_1+b_{n,m-1}(j))+a_{n,m}(j)b_{n,m-1}(j)h_{n,m}(j),
\end{multline}
where $k_{n,m}(j)=\frac{c_{n,m}(j)}{a_{n-1,m}(j)b_{n,m}(j)}-1$ and
$h_{n,m}(j)=\frac{c_{n,m}(j)}{a_{n,m}(j)b_{n,m-1}(j)}-1$ are {\it the Laplace invariants}
of the difference operators ${\cal L}_j$.

Suppose any two neighboring operators ${\cal L}_j$ and ${\cal L}_{j+1}$ are related by a {\it Darboux--Laplace transformation}, that is, they satisfy
the operator relation
$$
{\cal L}_{j+1}{\cal D}_j={\cal D}_{j+1}{\cal L}_j,
$$
where ${\cal D}_j=T_1+b_{n,m-1}(j)$. This relation is equivalent to the system of difference equations for the coefficients of the operators that
yields the following equation on the Laplace invariants:
\begin{equation}
\label{dht}
\frac{h_{n,m+1}(j)h_{n-1,m}(j)}{h_{n,m}(j)h_{n-1,m+1}(j)}=\frac{(1+h_{n,m}(j+1))(1+h_{n-1,m+1}(j-1))}{(1+h_{n,m}(j))(1+h_{n-1,m+1}(j))}.
\end{equation}
In terms of the new set of variables $h_{n,m}(j)=\exp(q_{n+1,m-1}(j+1)-q_{n,m}(j))$ this equation can be rewritten as follows:
\begin{equation}
\label{dqt}
\exp(q_{n+1,m+1}(j)+q_{n,m}(j)-q_{n+1,m}(j)-q_{n,m+1}(j))=\frac{1+\exp(q_{n+1,m}(j+1)-q_{n,m+1}(j))}{1+\exp(q_{n+1,m}(j)-q_{n,m+1}(j-1))}.
\end{equation}
The substitution $q_{n,m}(j)=u_{n,m}(j)-u_{n,m}(j-1)$ gives the following lattice:
\begin{multline}
\label{dut}
\exp(u_{n+1,m+1}(j)+u_{n,m}(j)-u_{n+1,m}(j)-u_{n,m+1}(j))=\\
=1+\exp(u_{n,m+1}(j-1)-u_{n,m+1}(j)-u_{n+1,m}(j)+u_{n+1,m}(j+1)).
\end{multline}
The equations~(\ref{dht})--(\ref{dut}) represent various forms of the discrete two-dimensional Toda lattice. The lattice~(\ref{dut}) is a
particular case of {\it discrete exponential systems} that were introduced in~\cite{GHY}:
\begin{multline}
\nonumber
\exp(u_{n+1,m+1}(j)+u_{n,m}(j)-u_{n+1,m}(j)-u_{n,m+1}(j))=\\
=1+\exp\left(\sum\limits_{k=1}^{j-1} a_{jk} u_{n,m+1}(k)+\frac{a_{jj}}{2}(u_{n,m+1} (j)+u_{n+1,m} (j))+
\sum\limits_{k=j+1}^{r} a_{jk} u_{n+1,m}(k)\right),
\end{multline}
where $j=1,2,\dots,r$. These systems are purely discrete analogs of the systems~(\ref{expsyst}),(\ref{dexpsyst}).
\begin{remark}
\rm
Purely discrete two-dimensional Toda lattice~(\ref{dut}) is a particular case of the well-known Hirota equation, see~\cite{Za}.
\end{remark}

Purely discrete analog of the lattice corresponding to the series $A$ is obtained if one imposes the constrains $h_{n,m}(-1)=h_{n,m}(r)=0$. These
constraints can be rewritten in terms of the variable $q_{n,m}$ as follows:
\begin{equation}
\label{prr}
q_{n,m}(-1)=\infty,\quad q_{n,m}(r+1)=-\infty.
\end{equation}
In purely discrete case, as well as in the semidiscrete one, the lattice~(\ref{dht}) admits reflections only about integer points. In terms of the
Laplace invariants such involution takes the form
$$
h_{n,m}(-j)=h_{n+j,m-j}(j).
$$
(here we consider reflection about zero). In terms of the variable $q_{n,m}$ for $j=1$ this involution can be rewritten as follows:
\begin{equation}
\label{boundc}
q_{n+1,m-1}(0)-q_{n,m}(-1)=q_{n+2,m-2}(2)-q_{n+1,m-1}(1).
\end{equation}
Discrete lattice that satisfies this constraint of the left edge and trivial boundary condition on the right edge is a purely discrete analog of the
$C$-series lattice. It tends to this lattice in the continuum limit and it is a reduction of the $A$-series lattice of the double length. Boundary
condition~(\ref{boundc}) was found in~\cite{Ha06}.
\begin{remark}
\rm
It is rather natural to expect that in this case nonlocal variables could be used to express symmetries and integrals along characteristics as well
as in the continuous and the semidiscrete ones. However in fact attempts to construct nonlocal variables meet certain obstacles. Another method of
obtaining symmetries for discrete exponential systems of small order, corresponding to simple Lie algebras was used in the paper~\cite{GHY}.
\end{remark}

The structure of two-dimensional Toda lattices connected with a sequence of Darboux--Laplace transformations allows to obtain a complete family
of independent $m$-integrals and $n$-integrals in the purely discrete case also. The following theorem is proved exactly in the same way as its
analogs in the other cases we have considered.
\begin{theorem}
The lattice~(\ref{dqt}) with trivial boundary conditions~(\ref{prr}) is Darboux integrable. The coefficients of the difference operator
\begin{equation}
\nonumber
{\mathcal A}=(T_1 +\exp(q_{n,m}(0)-q_{n+1,m}(0)))(T_1+\exp(q_{n,m}(1)-q_{n+1,m}(1)))\dots (T_1+\exp(q_{n,m}(r)-q_{n+1,m}(r)))
\end{equation}
form a complete family of independent $m$-integrals for this system. These $m$-integrals are defined explicitly by the formula
\begin{multline}
\label{pss}
J_k=\sum\limits_{0\leqslant i_1<i_2<\dots<i_k\leqslant r}^{}\exp (c_{n,m}(0)+\dots+c_{n,m} (i_1-1)+c_{n+1,m}(i_1+1)+\dots+c_{n+1,m}(i_2-1)+\\
+c_{n+2,m}(i_2+1)+\dots+c_{n+k-1,m}(i_r-1)+c_{n+k,m}(i_k+1)+\dots+c_{n+k,m}(r)),
\end{multline}
where $c_{n,m}(j)=q_{n,m}(j)-q_{n+1,m}(j)$.
\end{theorem}
\begin{remark}
\rm
Conversely to the continuous case, in the purely discrete case the variables $n$ and $m$ enter the equation~(\ref{dqt}) not very symmetrically. Therefore in
order to obtain explicit expressions for $n$-integrals one has not only to change the roles of the variables $n$ and $m$ in the formula~(\ref{pss}) but also has to
make an additional change of variables $q_{n,m}(j)\rightarrow -q_{n,m}(-j)$.
\end{remark}
\begin{example}
\rm
Let us give complete families of independent integrals in both directions explicitly for the purely discrete lattice corresponding to the series $A$ for $r=3$:
\begin{eqnarray}
\nonumber
J_0&=&\exp(c_{n,m}(0)+c_{n,m}(1)+c_{n,m}(2)+c_{n,m}(3)),\\
\nonumber
J_1&=&\exp(c_{n,m}(0)+c_{n,m}(1)+c_{n,m}(2))+\exp(c_{n,m}(0)+c_{n,m}(1)+c_{n+1,m}(3))+\\
\nonumber
&+&\exp(c_{n,m}(0)+c_{n+1,m}(2)+c_{n+1,m}(3))+\exp(c_{n+1,m}(1)+c_{n+1,m}(2)+c_{n+1,m}(3)),\\
\nonumber
J_2&=&\exp(c_{n,m}(0)+c_{n,m}(1))+\exp(c_{n,m}(0)+c_{n+1,m}(2))+\exp(c_{n,m}(0)+c_{n+2,m}(3))+\\
\nonumber
&+&\exp(c_{n+1,m}(1)+c_{n+1,m}(2))+\exp(c_{n+1,m}(1)+c_{n+2,m}(3))+\exp(c_{n+2,m}(2)+c_{n+2,m}(3)),\\
\nonumber
J_3&=&\exp(c_{n,m}(0))+\exp(c_{n+1,m}(1))+\exp(c_{n+2,m}(2))+\exp(c_{n+3,m}(3)),\\
\nonumber
I_0&=&\exp(-b_{n,m}(3)+b_{n,m}(2)+b_{n,m}(1)+b_{n,m}(0)),\\
\nonumber
I_1&=&\exp(-b_{n,m}(3)-b_{n,m}(2)-b_{n,m}(1))+\exp(-b_{n,m}(3)-b_{n,m}(2)-b_{n,m+1}(0))+\\
\nonumber
&+&\exp(-b_{n,m}(3)-b_{n,m+1}(1)-b_{n,m+1}(0))+\exp(-b_{n,m+1}(2)-b_{n,m+1}(1)-b_{n,m}(0)),\\
\nonumber
I_2&=&\exp(-b_{n,m}(3)-b_{n,m}(2))+\exp(-b_{n,m}(3)-b_{n,m+1}(1))+\exp(-b_{n,m}(3)-b_{n,m+2}(0))+\\
\nonumber
&+&\exp(-b_{n,m+1}(2)-b_{n,m+1}(1))+\exp(-b_{n,m+1}(2)-b_{n,m+2}(0))+\exp(-b_{n,m+2}(1)-b_{n,m+2}(0)),\\
\nonumber
I_3&=&\exp(-b_{n,m}(3))+\exp(-b_{n,m+1}(2))+\exp(-b_{n,m+2}(1))+\exp(-b_{n,m+3}(0)),
\end{eqnarray}
where $b_{n,m}(j)=q_{n,m}(j)-q_{n,m+1}(j)$.
\end{example}
\begin{corollary}
Purely discrete two-dimensional Toda lattices corresponding to the series $C$ are Darboux integrable.
\end{corollary}
\begin{remark}
\rm
The proof of this corollary follows from the fact that $C$-series lattices are reductions of $A$-series lattices exactly in the same way as in the two
previous cases. However we can not use this idea to prove Darboux integrability of discrete exponential systems corresponding to Lie algebras of the series
$B$ and $D$ because the question of whether these lattices are reductions of an $A$-series lattice still remains an open problem.
\end{remark}
\begin{remark}
\rm
The results obtained here go along with the general ideology of the papers~\cite{HZY,GHY}. Roughly speaking this ideology is based on the principle that
proper discretization should leave symmetries and integrals along characteristics unchanged. Clearly, in our situation the same method of obtaining integrals
works in the continuous case, in the semidiscrete case and in the purely discrete case, and the explicit formulas for $x$-integrals of the semidiscrete
system~(\ref{todaq}) give $m$-integrals for its complete discretization~(\ref{dqt}).
\end{remark}

\section{Acknowledgements}

In conclusion I'd like to thank I.~T.~Habibullin for useful discussions and for valuable remarks. This research was partially supported by the Russian
President grant HSh-4833.2014.1, by the Russian Government grant 2010-220-01-077 and by RFBR grant no.~14-01-00012-a.

\end{document}